\documentclass[aps,prd,nofootinbib,twocolumn,10pt]{revtex4}
\usepackage{amsmath}
\usepackage{amsfonts}
\usepackage{txfonts}
\usepackage{graphicx}
\usepackage{dcolumn}
\usepackage{bbm}
\usepackage{amssymb}
\usepackage{latexsym}
\usepackage{CJK}
\usepackage{titlesec}
\usepackage[colorlinks=true, linkcolor=red, citecolor=blue]{hyperref}
\begin{document}
 \bibliographystyle{unsrt}

\title{Comparing holographic dark energy models with statefinder}
\author{Jing-Lei Cui, Jing-Fei Zhang\footnote{Corresponding author}}
\email{jfzhang@mail.neu.edu.cn}
\affiliation{Department of Physics, College of Sciences, Northeastern University,
Shenyang 110004, China}

\begin{abstract}
We apply the statefinder diagnostic to the holographic dark energy models, including the original holographic dark energy (HDE) model, the new holographic dark energy model, the new agegraphic dark energy (NADE) model, and the Ricci dark energy model. In the low-redshift region the holographic dark energy models are degenerate with each other and with the $\Lambda$CDM model in the $H(z)$ and $q(z)$ evolutions. In particular, the HDE model is highly degenerate with the $\Lambda$CDM model, and in the HDE model the cases with different parameter values are also in strong degeneracy. Since the observational data are mainly within the low-redshift region, it is very important to break this low-redshift degeneracy in the $H(z)$ and $q(z)$ diagnostics by using some quantities with higher order derivatives of the scale factor. It is shown that the statefinder diagnostic $r(z)$ is very useful in breaking the low-redshift degeneracies. By employing the statefinder diagnostic the holographic dark energy models can be differentiated efficiently in the low-redshift region. The degeneracy between the holographic dark energy models and the $\Lambda$CDM model can also be broken by this method. Especially for the HDE model, all the previous strong degeneracies appearing in the $H(z)$ and $q(z)$ diagnostics are broken effectively. But for the NADE model, the degeneracy between the cases with different parameter values cannot be broken, even though the statefinder diagnostic is used. A direct comparison of the holographic dark energy models in the $r$--$s$ plane is also made, in which the separations between the models (including the $\Lambda$CDM model) can be directly measured in the light of the current values $\{r_0,~s_0\}$ of the models.
\end{abstract}
\maketitle

\renewcommand{\thesection}{\arabic{section}}
\renewcommand{\thesubsection}{\arabic{subsection}}
\titleformat*{\section}{\flushleft\bf}
\titleformat*{\subsection}{\flushleft\bf}
\section {Introduction}
The observations from type Ia supernovae (SNIa)~\cite{Riess:1998cb}, large scale structure (LSS)~\cite{Tegmark:2003ud}, and cosmic microwave background (CMB) anisotropies~\cite{Spergel:2003cb} have shown convincing evidence supporting the cosmic acceleration. This cosmic acceleration is attributed to a mysterious dominant component, dark energy (DE), with negative pressure. The combined analysis of cosmological observations suggests that the universe is spatially flat, and that it consists of about $73\%$ DE, $27\%$ dust matter, and negligible radiation. Nowadays physicists have already constructed numerous models of DE, though its nature still remains enigmatic. The simplest DE model is the $\Lambda$CDM cosmology, in which DE has the equation of state $w=-1$. However, the cosmological constant scenario has to face the so-called ``fine-tuning problem''
and ``coincidence problem''~\cite{lamada}. In light of scalar fields, some dynamical DE models have been studied widely, in which the equation of state parameter $w$ is dependent on time, for example, quintessence~\cite{quintessence}, phantom~\cite{Caldwell:1999ew}, quintom~\cite{Feng:2004ad}, tachyon~\cite{Sen:2002in}, ghost condensate~\cite{ArkaniHamed:2003uy} models, and so on.

The cosmological constant problem is essentially an issue of quantum gravity, because of the significance of the vacuum expectation value of some quantum fields in the cosmology (gravity) context. However, we have no full theory of quantum gravity yet. But theorists have been making lots of efforts to try to resolve the cosmological constant problem by various means. In recent years, based on the holographic principle, a series of the holographic DE models were proposed. Based on some considerations of the holographic principle, the total energy in a region of size $L$ should not exceed the mass of a black hole of the same size, thus $L^3\rho _{de}\lesssim LM^2_p$~\cite{Cohen:1998zx}. The largest $L$ allowed is the one saturating this inequality, thus the dark energy density is $\rho _{de} = 3c^2M^2_pL^{-2}$, where $c$ is an introduced numerical constant characterizing some uncertainties in the effective quantum field theory, $M_p$ is the reduced Planck mass, defined by $M^2_p = (8\pi G)^{-1}$, and $L$ is the infrared (IR) cutoff in the theory. In the holographic DE model~\cite{Li:2004rb}, Li proposed that the IR cutoff $L$ should be given by the future event horizon of the universe. Hereafter, different DE models of this holographic type have been proposed, such as the new holographic DE model~\cite{Li:2012xf}, the new agegraphic DE model~\cite{Wei:2007ty}, and the Ricci DE model~\cite{Gao:2007ep}, which can be uniformly  called holographic DE models for simplicity.  All these holographic DE models can be used to interpret or describe the cosmic acceleration~\cite{hde1,hde2}.

As the amount of DE models is increasing, how to discriminate DE models becomes an important problem. One may consider the use of the Hubble parameter $H=\dot{a}/a$ and the deceleration parameter $q\equiv-\ddot{a}/(aH^2)$ to differentiate DE models, where $a(t)$ is the scale factor of a Friedmann-Robertson-Walker (FRW) universe. $H$ and $q$ containing, respectively, first-derivative and second-derivative of $a(t)$, are geometrical variables characterizing the cosmic expansion. However, if we want to diagnose some similar DE models or one DE model with different values of parameters, perhaps we should employ higher order derivatives of $a(t)$. Sahni et al.~\cite{Sahni:2002fz} introduced the statefinder diagnostic $\{r,~s\}$, in which derivatives of $a(t)$ are up to the third order, to do this job. This distinguishing method is a geometrical diagnosis in a model-independent manner and in principle needs high-precision observational data. Since different DE models exhibit different evolution trajectories in the $r$--$s$ plane, especially being separated distinctively with the values of $\{r_0,~s_0\}$, the statefinder can be used to diagnose different DE models~\cite{sfide,sfhde,ngcg,Zhang:2007uh}.

In this paper, we use the statefinder parameters $\{r,s\}$ to diagnose holographic DE models including the original holographic dark energy (HDE) model, the new holographic dark energy (NHDE) model, the new agegraphic dark energy (NADE) model, and the Ricci dark energy (RDE) model. In Sect. 2, a series of the holographic DE models are briefly reviewed. Diagnosing holographic DE models with $H$, $q$, and the statefinder is discussed in Sect. 3. In Sect. 4, the conclusion is given.

\section {Holographic dark energy models}
\subsection*{2.1 The HDE model}
We consider a spatially flat FRW universe containing dark energy and matter (note that we assume a flat universe in the whole paper), the Friedmann equation is
\begin{equation}
H^2=\frac{1 }{3M^2_p}(\rho_{de}+\rho_m),\label{eq1}
\end {equation}
where $\rho_m$ and $\rho_{de}$ are, respectively, energy densities for matter and dark energy.

In the HDE model~\cite{Li:2004rb}, $\rho _{de} = 3c^2M^2_pL^{-2}$, and $L$ is the future event horizon given by
\begin{equation}
L=a\int\limits_a^\infty\frac{da'}{Ha'^2}.\label{eq2}
\end{equation}
By using Eqs. (\ref{eq1}) and (\ref{eq2}), we can obtain the equation for the fractional density of the
dark energy ($ \Omega_{de}\equiv \rho_{de}/{3M^2_p}H^2)$,
\begin{equation}
\Omega '_{de}= \Omega_{de}(1- \Omega_{de})\left(1+\frac{2}{c}\sqrt{ \Omega_{de}}\right),\label{eq3}
\end{equation}
where the prime denotes the derivative with respect to $x=\ln a$. Then from the energy conservation equation,  $\dot{\rho}_{de}+3H(1+w)\rho_{de}=0$, the equation of state (EOS) of HDE, $w\equiv p_{de}/\rho_{de}$, can be given:
\begin{equation}
w = -\frac{1}{3}-\frac{2}{3c}\sqrt{ \Omega_{de}}.\label{eq4}
\end{equation}

\subsection*{2.2 The NHDE model}
Recently, a new holographic dark energy model with the action principle was proposed~\cite{Li:2012xf}, in which the dark energy density reads
\begin{equation}
\rho_{de}=M^2_p\left(\frac{d}{a^2L^2}+\frac{\lambda}{2a^4}\right),\label{eq17}
\end{equation}
where $d$ is a numerical parameter, and
\begin{equation}
L=\int\limits_t^\infty\frac{dt'}{a(t')}+L(a=\infty),~~~\dot{L}=-\frac{1}{a}\label{eq18},
\end{equation}
\begin{equation}
\lambda=\int\limits_0^t\frac{4a(t')ddt'}{L^3(t')}+\lambda(a=0),~~~\dot{\lambda}=-\frac{4ad}{L^3},\label{eq19}
\end{equation}
where the dot denotes a derivative with respect to time $t$. As proven in ~\cite{Li:2012xf}, the equations of motion force $L (a=\infty)=0$, so $aL$ is exactly the future event horizon. The $\lambda (a=0)$ term behaves in the same way as radiation; thus, it can be naturally interpreted as dark radiation~\cite{Hamann:2010bk}.
For convenience, we define the ``Hubble-free" quantities
\begin{equation}
\tilde{L}\equiv H_0L,~~\tilde{\lambda}\equiv \frac{\lambda}{H^2_0},~~E=\frac{H}{H_0}.\label{eq20}
\end{equation}
Combining Eqs. (\ref{eq17})--(\ref{eq20}) with the Friedmann equation, we get
\begin{equation}
E=\sqrt{\Omega_{m0}a^{-3}+\frac{1}{3}\left(\frac{da^{-2}}{\tilde{L}^2}+\frac{\tilde{\lambda}a^{-4}}{2}\right)},\label{eq21}
\end{equation}
where $\Omega_{m0}=\rho_{m0} /\rho_0$ is the fractional density of matter in the present universe ($\rho_0 =3M^2_p H^2_0$ is today's critical density of the universe). From Eq. (\ref{eq21}), we get the fractional density of the dark energy $\Omega_{de}$,
\begin{equation}
\Omega_{de}=-\frac{1}{3E^2}\left(\frac{da^{-2}}{\tilde{L}^2}+\frac{\tilde{\lambda}a^{-4}}{2}\right).\label{eq22}
\end{equation}
Furthermore, by using Eqs. (\ref{eq1}), (\ref{eq17}) and the conservation equation of energy, the EOS of NHDE takes the form
\begin{equation}
w=\frac{\tilde{\lambda}\tilde{L}^2-2da^2}{3\tilde{\lambda}\tilde{L}^2+6da^2}. \label{eq23}
\end{equation}

\subsection*{2.3 The NADE model}
In the NADE model~\cite{Wei:2007ty},  $\rho _{de} = 3n^2M^2_p\eta ^{-2}$, where $n$ is the introduced numerical parameter, and the conformal time $\eta$ is written as
\begin{equation}
\eta =\int\limits_0^a\frac{da'}{Ha'^2}.\label{eq5}
\end{equation}
By using Eqs. (\ref{eq1}), (\ref{eq5}), and the energy conservation equation, we can obtain the equation of motion for $ \Omega_{de}$ and the EOS of NHDE,
\begin{equation}
\Omega '_{de}= \Omega_{de}(1- \Omega_{de})\left(3-\frac{2}{na}\sqrt{ \Omega_{de}}\right),\label{eq6}
\end{equation}
\begin{equation}
w = -1+\frac{2}{3na}\sqrt{ \Omega_{de}}.\label{eq7}
\end{equation}

\subsection*{2.4 The RDE model}

In the RDE model~\cite{Gao:2007ep}, $L$ in the definition of holographic type dark energy $\rho _{de} = 3c^2M^2_pL^{-2}$ is connected to the Ricci scalar curvature,
\begin{equation}
\mathcal{R}=-6(\stackrel{\centerdot}{H}+2H^2)\label{eq8},
\end{equation}
where the dot denotes a derivative with respect to time $t$. So the Ricci dark energy is
\begin{equation}
\rho_{de}=3\alpha M^2_p(\stackrel{\centerdot}{H}+2H^2),\label{eq9}
\end{equation}
where $\alpha $ is a dimensionless coefficient. The Friedmann equation can be rewritten as
\begin{equation}
E^2=\frac{H^2}{H_0^2}=\Omega_{m0}e^{-3x}+\alpha\left(\frac{1}{2}\frac{dE^2}{dx}+2E^2\right),
\end{equation}
where $x=\ln{a}$. Solving the Friedmann equation we get the result
\begin{equation}
E^2=\Omega_{m0}e^{-3x}+\frac{\alpha}{2-\alpha}\Omega_{m0}e^{-3x}+f_0e^{-(4-\frac{2}{\alpha})x},\label{eq18}
\end{equation}
where $f_0$ is an integration constant. Using the initial condition $E_0 = 1$, the integration constant $f_0$ is determined as
\begin{equation}
f_0=1-\frac{2}{2-\alpha}\Omega_{m0}.
\end{equation}
In Eq. (\ref{eq18}), the last two terms on the right-hand side is the contribution of the dark energy $\rho_{de}/\rho_0$. So the fractional density of the dark energy is
\begin{equation}
\Omega_{de}=\frac{1}{E^2}\frac{\rho_{de}}{\rho_0}=\frac{1}{E^2}\left(\frac{\alpha}{2-\alpha}\Omega_{m0}e^{-3x}+f_0e^{-(4-\frac{2}{\alpha})x}\right).\label{eq10}
\end{equation}
Furthermore, from the energy conservation equation, we can obtain the EOS of RDE
\begin{equation}
w=\frac{\frac{\alpha-2}{3\alpha}f_0e^{-(4-\frac{2}{\alpha})x}}{\frac{\alpha}{2-\alpha}\Omega_{m0}e^{-3x}+f_0e^{-(4-\frac{2}{\alpha})x}}.
\end{equation}

\section {Statefinder diagnostic}

\begin{figure*}[htbp]
\centering
\includegraphics[scale=0.6]{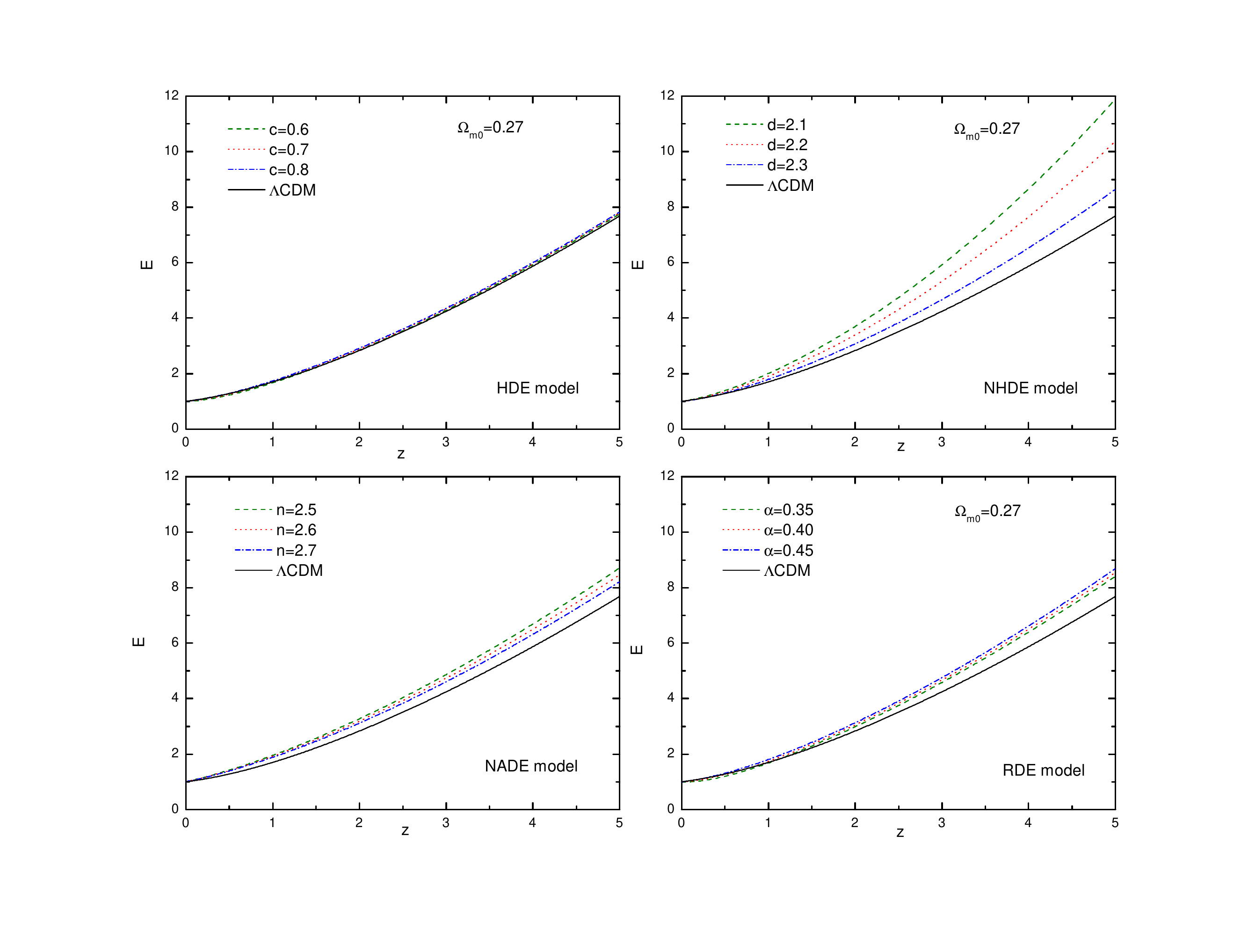}
\caption{\label{fig1} Evolutions of the dimensionless Hubble parameter $E$ with redshift $z$ for the HDE, NHDE, NADE, and RDE models. The $E(z)$ curve of the $\Lambda$CDM model is also shown for a comparison}
\end{figure*}
\begin{figure}[htbp]
\centering
\includegraphics[scale=0.25]{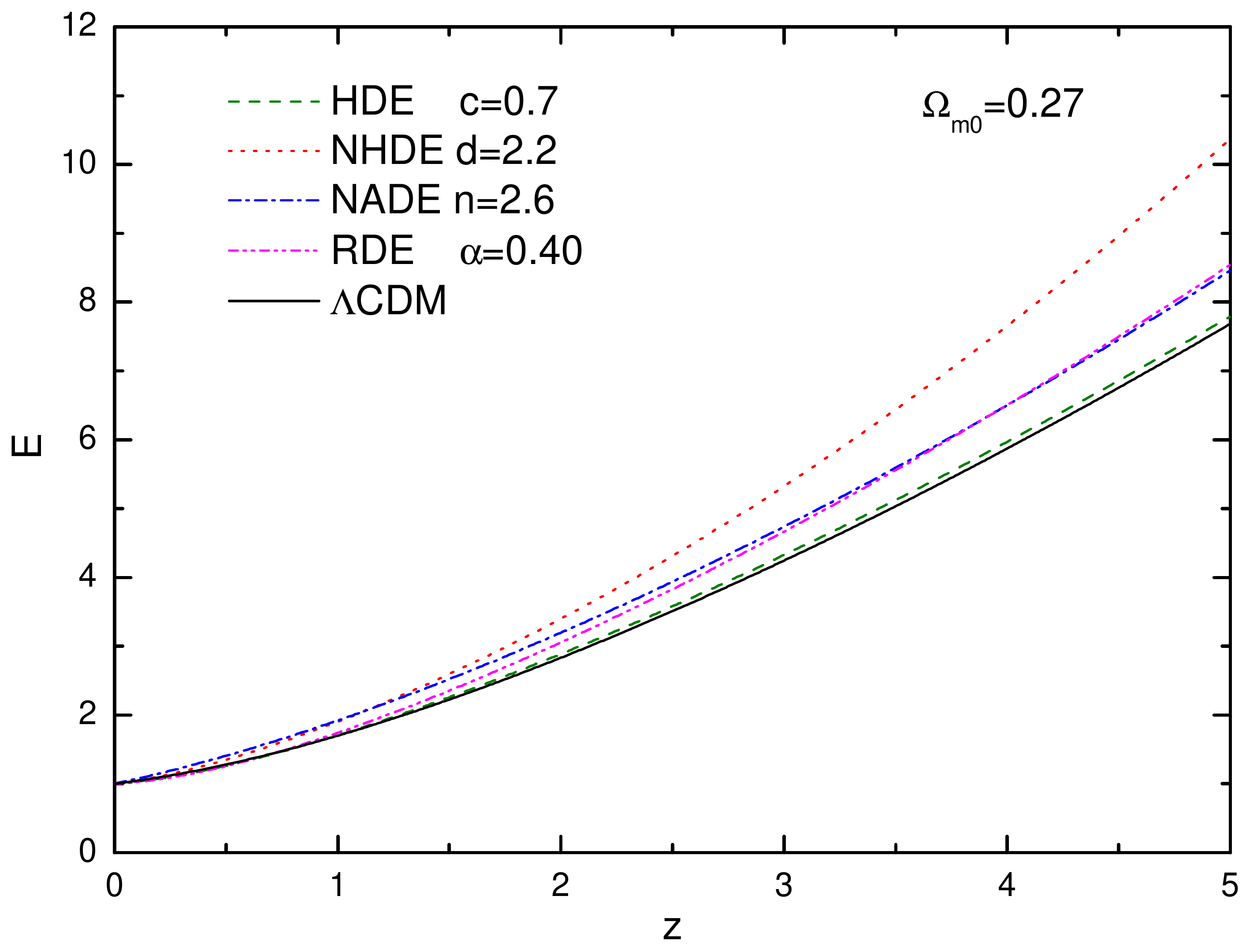}
\caption{\label{fig2} Comparison of the holographic DE models in the $E(z)$ evolution diagram. The $\Lambda$CDM model is also shown for a comparison}
\end{figure}

\begin{figure*}[htbp]
\centering
\includegraphics[scale=0.5]{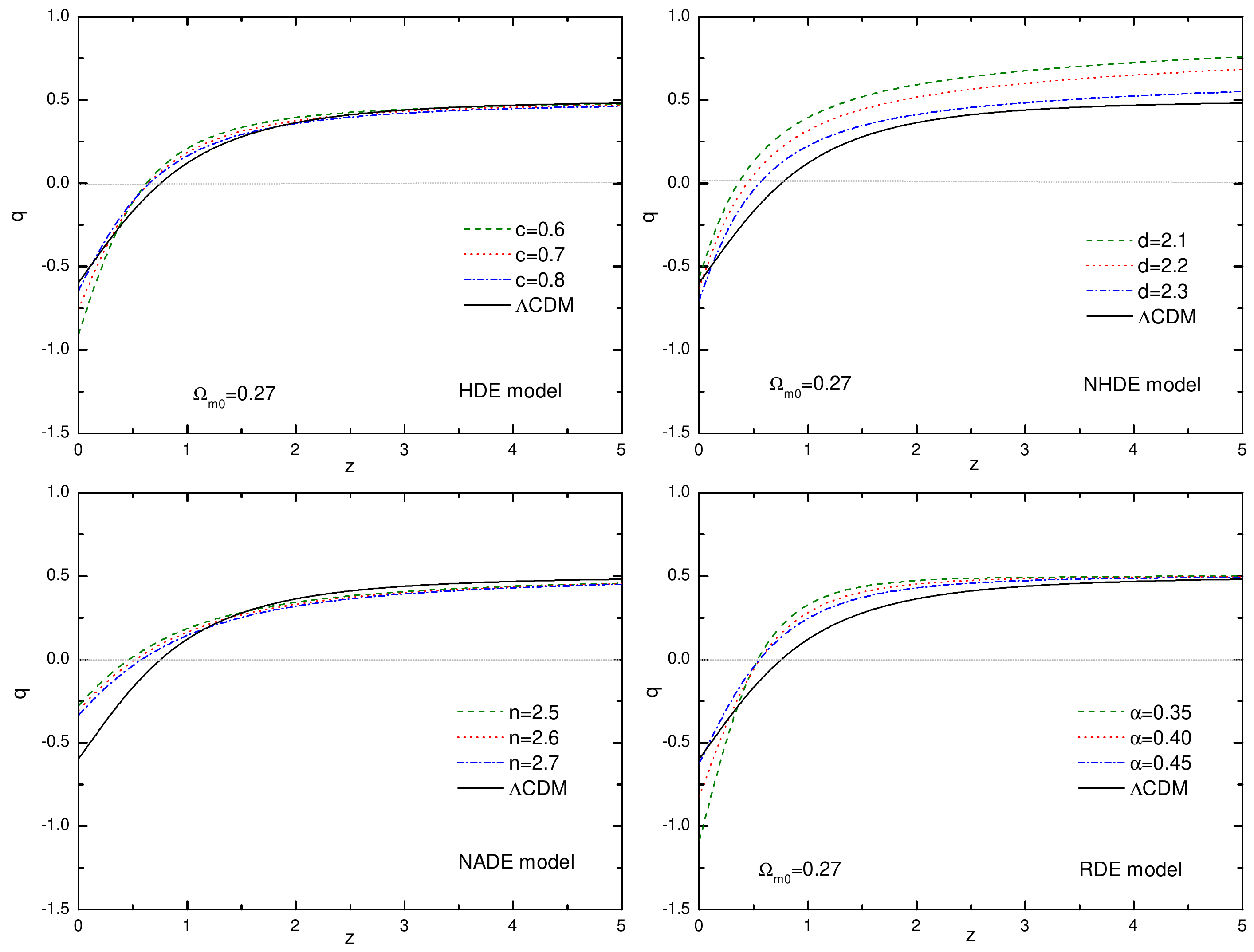}
\caption{\label{fig3} Evolutions of the deceleration parameter $q$ with redshift $z$ for the HDE, NHDE, NADE, and RDE models. The $q(z)$ curve of the $\Lambda$CDM model is also shown for a comparison}
\end{figure*}
\begin{figure}[htbp]
\centering
\includegraphics[scale=0.25]{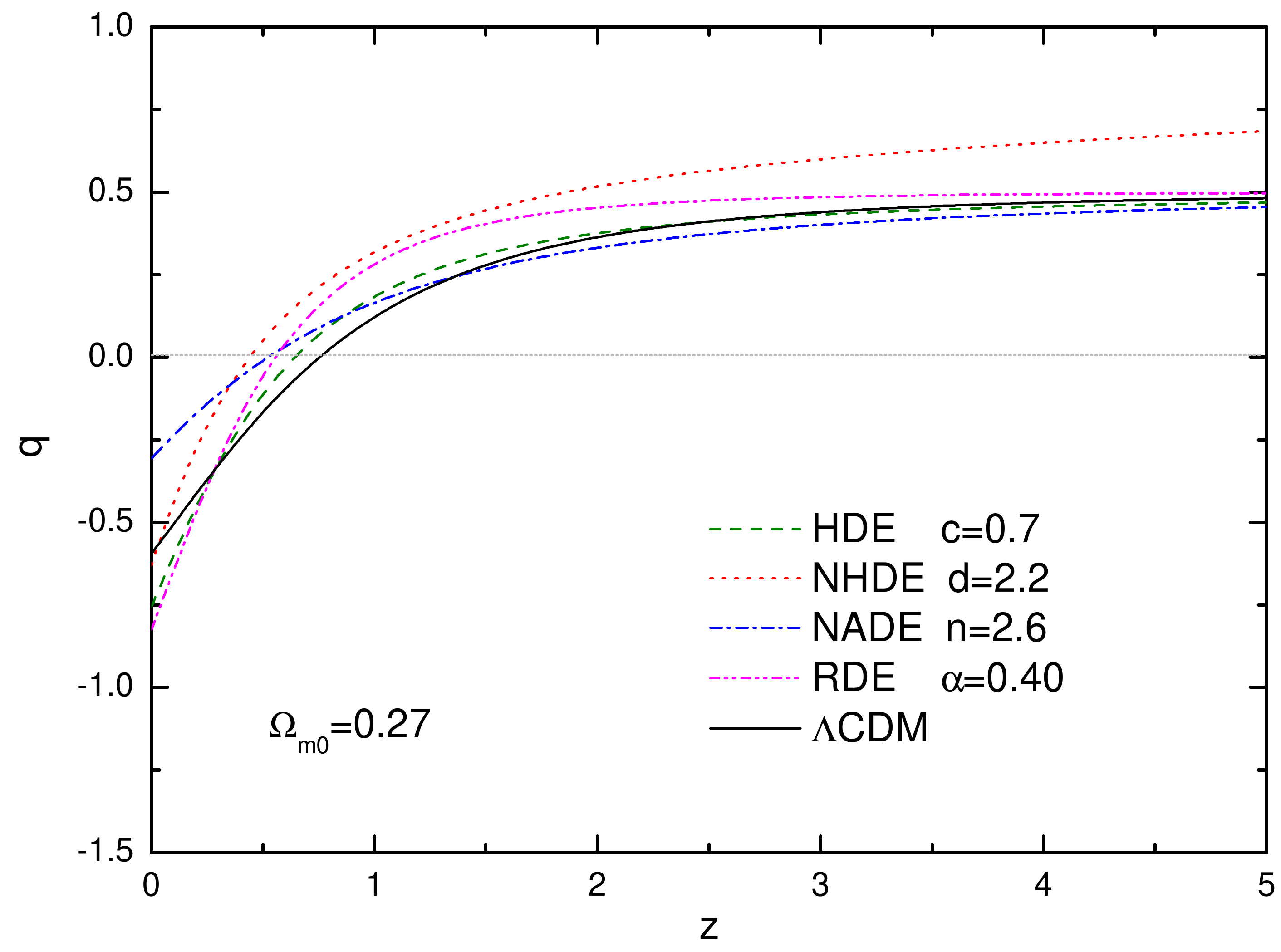}
\caption{\label{fig4} Comparison of the holographic DE models in the $q(z)$ evolution diagram. The $\Lambda$CDM model is also shown for a comparison}
\end{figure}

\begin{figure*}[htbp]
\centering
\includegraphics[scale=0.5]{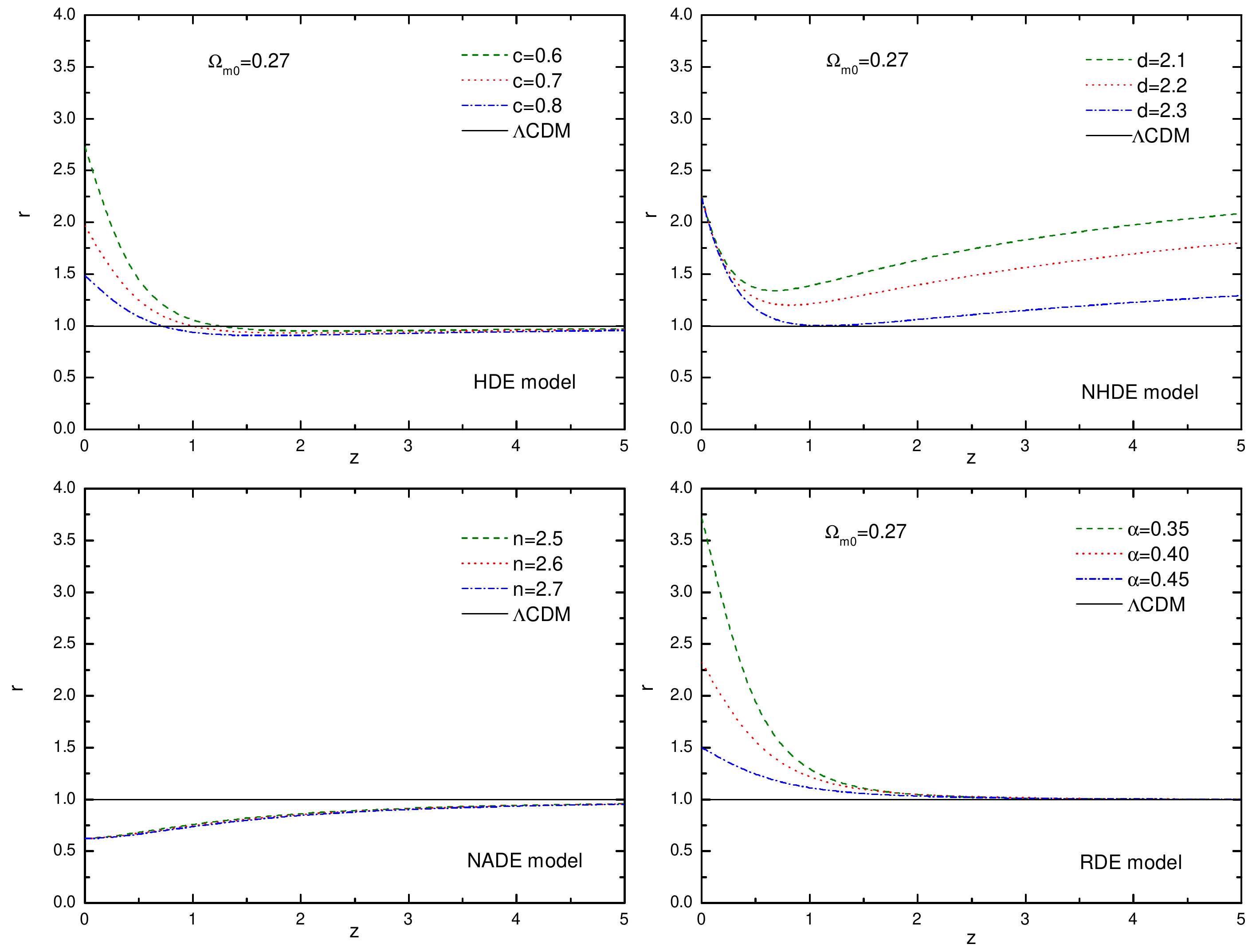}
\caption{\label{fig5} Evolutions of the statefinder parameter $r$ with redshift $z$ for the HDE, NHDE, NADE, and RDE models. The line $r=1$ indicates the $\Lambda$CDM mode}
\end{figure*}
\begin{figure}[htbp]
\centering
\includegraphics[scale=0.25]{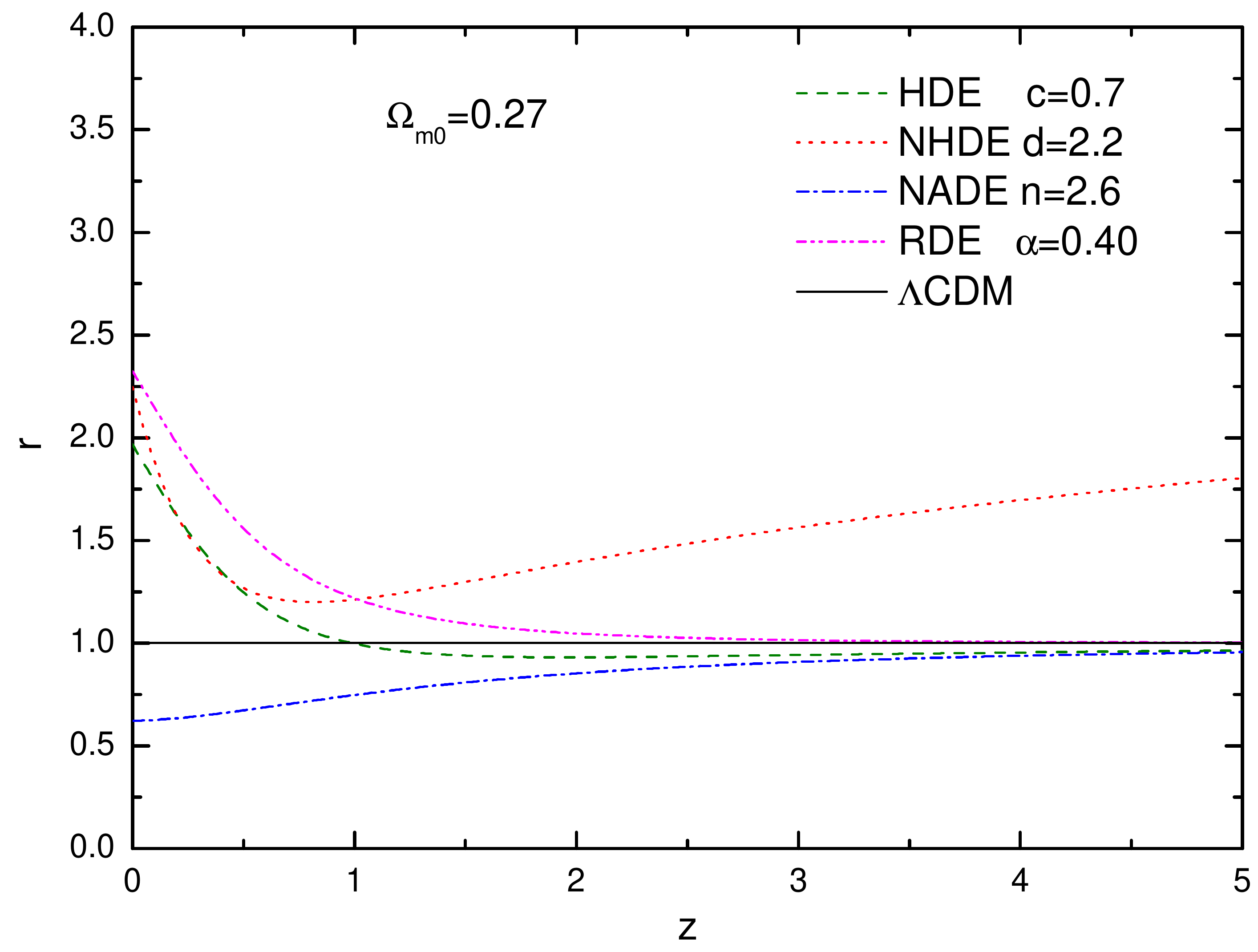}
\caption{\label{fig6} Comparison of the holographic DE models in the $r(z)$ evolution diagram. The $\Lambda$CDM model is also shown for a comparison}
\end{figure}

\begin{figure*}[htbp]
\centering
\includegraphics[scale=0.5]{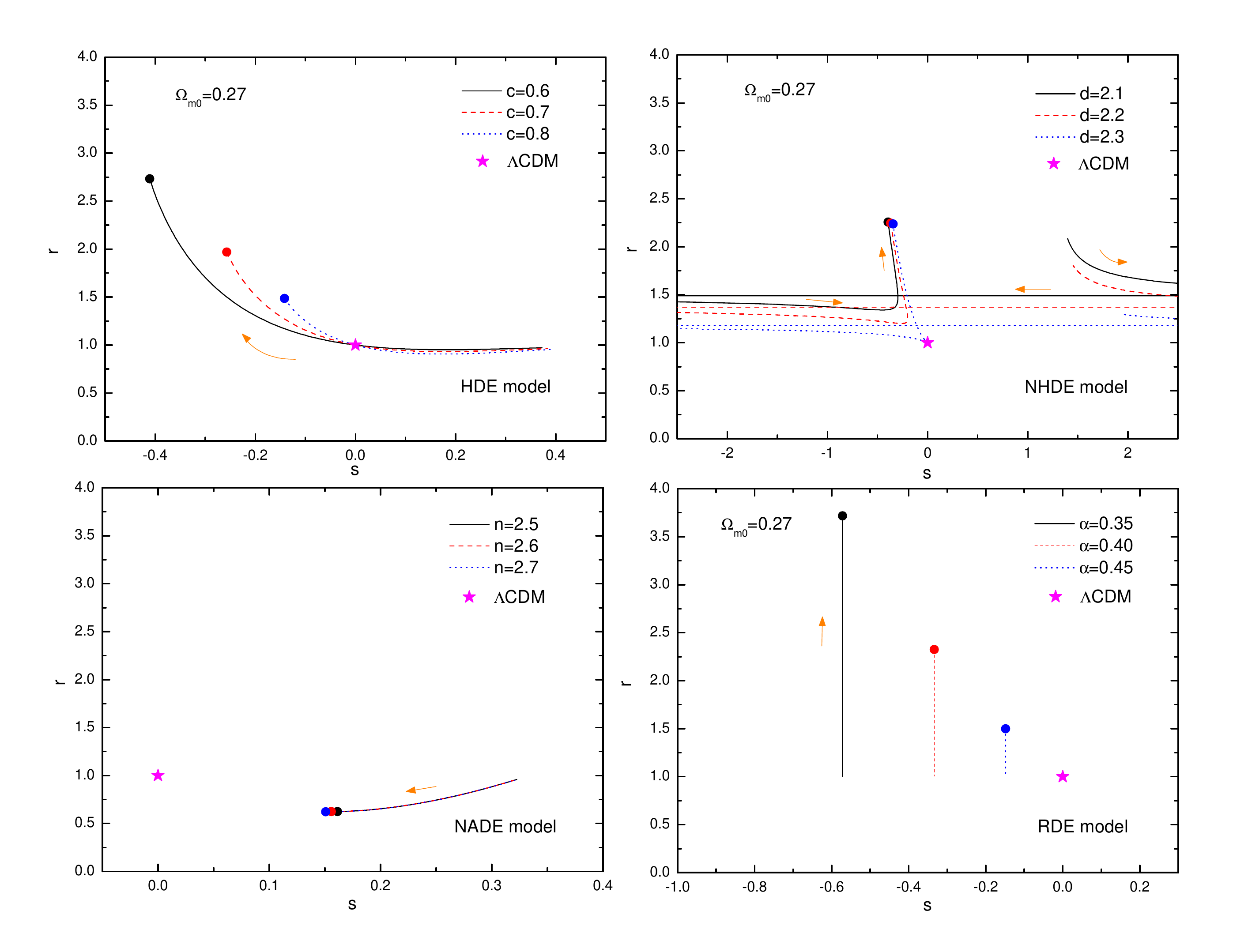}
\caption{\label{fig7} The evolutionary trajectories of $r(s)$ for the HDE, NHDE, NADE, and RDE models in the $r$--$s$ plane. The today's values of the statefinder pair $\{r_0,~s_0\}$ of the holographic DE models are marked by the round dots. The $\Lambda$CDM model is a fixed point (0,~1) in this plot, marked by a star. The arrows indicate the evolution directions}
\end{figure*}
\begin{figure}[htbp]
\centering
\includegraphics[scale=0.25]{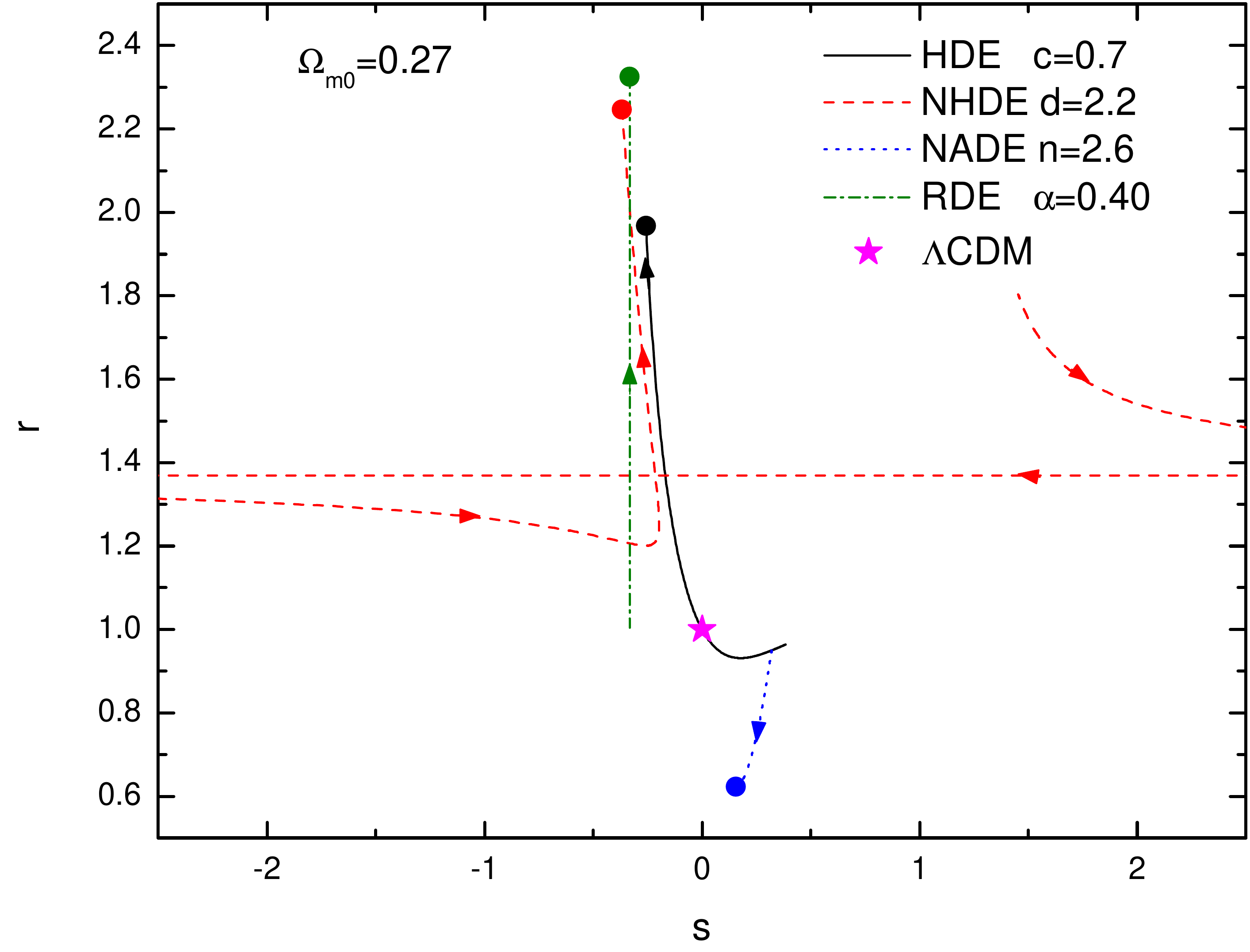}
\caption{\label{fig8} Comparison of the evolutionary trajectories $r(s)$ of the holographic DE models in the $r$--$s$ plane. The $\Lambda$CDM model, denoted by a star, is also shown for a comparison. The arrows indicate the evolution directions of the models}
\end{figure}

Before the presence of the statefinder diagnostic pair $\{r,s\}$, one can discriminate different dark energy models by only employing the evolutionary behaviors of the Hubble expansion rate $E$ and the deceleration parameter $q$. Note that knowledge of the expansion rate, $E(z)$, allows one to determine $Om$, which can then be used
as a null test for the cosmological constant~\cite{om}.
So, in this work, we also consider the evolutions of $E$ and $q$ of the above four holographic DE models.

Firstly, we consider the evolutions of $E$ for the models. In the HDE and NADE models, $E(z)$ can be expressed as
\begin{equation}
E=\frac{a^{-\frac{3}{2}}\sqrt{\Omega_{m0}}}{\sqrt{1-\Omega_{de}}}.
\end{equation}
For the NHDE and RDE models, the expressions of $E(z)$ are directly given by Eqs. (\ref{eq21}) and (\ref{eq18}), respectively.
The cosmological evolutions of the dark energy models can directly be given by the $E(z)$ diagrams.

In Fig.~\ref{fig1}, we plot the evolutions of the Hubble expansion rate $E$ with redshift $z$ for the HDE, NHDE, NADE, and RDE models,
and compare these dynamical models  with the $\Lambda$CDM model. We fix $\Omega_{m0}=0.27$ for all the models.
For properly choosing the typical values of the parameters in these models, we refer to the current observational constraints on the models.
In HDE, the parameter $c$ is taken to be 0.6, 0.7, and 0.8~\cite{Wang:2012uf}. In NHDE, the parameter $d$ is taken to be 2.1, 2.2, and 2.3~\cite{Li:2012fj}.
Since the NADE model is a single-parameter model, we apply the initial condition $\Omega_{de}(z_{ini})=n^2(1+z_{ini})^{-2}/4$ at $z_{ini}=2000$~\cite{Wei:2007xu}; the parameter $n$ is taken to be 2.5, 2.6, and 2.7~\cite{Zhang:2013lea}. In RDE, we choose $\alpha=0.35$, 0.40, and 0.45~\cite{Zhang:2009un}.

We can see from Fig.~\ref{fig1} that in the low-redshift region ($z\lesssim 1$) the difference between the holographic model and the $\Lambda$CDM model as well as that between different values of the parameter in one model cannot be effectively identified for all the cases, although the differences in the high-redshift region are rather evident. For HDE, all the $E(z)$ curves are nearly degenerate, for both the low-redshift and the high-redshift regions. Thus, obviously, the $E(z)$ diagnostic is not useful in differentiating the HDE model from the $\Lambda$CDM model as well as in discriminating parameter values in the HDE model. For the other three models, the situations are better, more or less, but in the low-redshift region the $E(z)$ curves are still nearly degenerate, and the differences can only be diagnosed in the high-redshift region. However, it is well known that the observational data are mainly within the low-redshift region, typically $z\lesssim 1$; for example, for the type Ia supernovae~\cite{Conley:2011ku}, the majority of the redshifts is in the range of $z<1$, and only a few of them are in the higher redshift range, $1<z<1.4$. Therefore, the $E(z)$ curves from the current observations cannot provide helpful diagnostics for the DE models. In order to see the degeneracy situation more clearly, we plot the $E(z)$ evolutions of the four holographic DE models and the $\Lambda$CDM model in Fig.~\ref{fig2}. From this figure, we can explicitly see that the DE models are in strong degeneracy in low-redshift region with the $E(z)$ diagnostic. If the Extremely Large Telescopes with high-resolution of the next generation come into use, they would observe the high-redshift QSOs ($2<z<5$)~\cite{Liske:2008ph}. Then perhaps the combination with the accurate high-redshift data can effectively differentiate DE models by using the $E(z)$ diagnostic.

Next, let us consider the situation as regards the deceleration parameter $q(z)$. For convenience, we express the deceleration parameter as
\begin{equation}
q=\frac{1}{2}+\frac{3}{2}w\Omega_{de}.\label{eq25}
\end{equation}
In Fig.~\ref{fig3}, we plot the $q(z)$ evolutions for the HDE, NHDE, NADE, and RDE models, also compared with the $\Lambda$CDM model. From this figure, we find that in some cases the degeneracy between the models in the $E(z)$ case is broken in some degree. For example, for the NHDE model the $q(z)$ curves  separate evidently, for the case between NHDE and $\Lambda$CDM and also for the case of NHDE with different parameter values. For the NADE model, we can see that by using the $q(z)$ diagnostic the NADE can be effectively differentiated from the $\Lambda$CDM in the low-redshift region, but the case of NADE with different parameter values is still in strong degeneracy during the whole evolution history. For the HDE and the RDE models, we find that the degeneracy situations are still severe, even though the $q(z)$ diagnostic is used. In particular, for the HDE case, we can discriminate neither the difference between the HDE and the $\Lambda$CDM nor the difference of the model with different parameter values. Likewise, we also plot the $q(z)$ evolutions of the four holographic DE models and the $\Lambda$CDM model in Fig.~\ref{fig4}. It is clear to see that effectively differentiating them with the $q(z)$ diagnostic is fairly difficult, if not impossible.

Therefore, we will consider the statefinder parameters $\{r,s\}$, defined by \cite{Sahni:2002fz}
\begin{equation}
r=\frac{\dddot{a}}{aH^3},
\end{equation}
and
\begin{equation}
s=\frac{r-1}{3(q-\tfrac{1}{2})}.
\end{equation}
For convenience, we use the following expressions, derived before:
\begin{equation}
r=1+\frac{9}{2}\Omega_{de}w(1+w)-\frac{3}{2}\Omega_{de}w',~~~s=1+w-\frac{w'}{3w},
\end{equation}
where the prime denotes the derivative with respect to $x=\ln a$.
Figure \ref{fig5} shows the evolutions of the statefinder parameter $r$ with $z$ in the four holographic DE models, also compared with the $\Lambda$CDM model. It is of interest to see that the cases are distinctively differentiated in the low-redshift region. For the HDE and the RDE models, we see that in the low-redshift region the $r(z)$ curves separate distinctively. From the above analysis we have learned that for the HDE model the cases with different parameter values are nearly degenerate with both $E(z)$ and $q(z)$ diagnostics; the HDE model is also degenerate with the $\Lambda$CDM model if these two diagnostics are used; however, we find that these degeneracies are effectively broken by using the statefinder diagnostic $r(z)$. For the NHDE model, the difference between the model and the $\Lambda$CDM can be clearly identified at $z\sim 0$, and the difference of the model with different parameter values can be effectively figured out within $z\sim 0.5-1$ with the statefinder diagnostic. For the NADE model, we find that the difference between the model and the $\Lambda$CDM can easily be distinguished, but the difference of the model with different $n$ values cannot be discriminated even though the statefinder parameter $r(z)$ is employed. Figure~\ref{fig6} compares the four holographic DE models and the $\Lambda$CDM model with the statefinder diagnostic $r(z)$; the differentiation of the models in the low-redshift region is directly seen from this figure.

Furthermore, we plot the evolution trajectories $r(s)$ of the four holographic DE models in Fig.~\ref{fig7}. The present values $\{r_0,~s_0\}$ of the holographic DE models are marked by the round dots; the $\Lambda$CDM model is a fixed point (0,~1) in this plot, marked by a star. In the $r$--$s$ plane, we can directly measure the difference of the models. The difference between the dynamical DE model and the $\Lambda$CDM model is measured by the separation of the round dot and the star; and the difference of the model with different parameter values can be measured by the separation of the dots. Since today's values of statefinder pair $\{r_0,~s_0\}$ are thought of as having been extracted from the low-redshift observational data, this measure implies the differentiation of DE models from the  low-redshift observational data. A direct comparison of the models in the $r$--$s$ plane is shown in Fig.~\ref{fig8}. If the accurate information of $\{r_0,~s_0\}$ can be extracted from the future high-precision observational data, one can discriminate DE models directly from the experiments with the statefinder diagnostic $\{r_0,~s_0\}$. If, furthermore, the precision high-redshift data can be obtained and combined with low-redshift data, one can even reconstruct the $r(s)$ trajectory to discriminate DE models and determine the property of DE.

\section {Conclusion}

There are several dynamical DE models originating from the consideration of the holographic principle. Therefore, it is of interest to discriminate these holographic DE models with the observational data. However, usually, these models are degenerate with each other in some degree, especially in the low-redshift region. In this paper, we analyze four typical holographic DE models, i.e., the HDE, NHDE, NADE, and RDE models, and we apply the statefinder diagnostic to discriminate them.

We have shown that in the low-redshift region the holographic DE models cannot be effectively discriminated with both the $E(z)$ and the $q(z)$ diagnostics. Also, the holographic DE models are nearly degenerate with the $\Lambda$CDM model in the low-redshift region if the $E(z)$ and $q(z)$ diagnostics are used. In particular, the HDE model is highly degenerate with the $\Lambda$CDM model, and in the HDE model the cases with different parameter values are also in strong degeneracy. Since the observational data are mainly within the low-redshift region (typically $z\lesssim 1$), it is rather important to break this low-redshift degeneracy appearing in the $E(z)$ and $q(z)$ diagnostics by using some quantities with higher order derivatives of the scale factor. We have shown that the statefinder diagnostic $r(z)$ is very helpful in doing this job.

We have applied the statefinder diagnostic to the holographic DE models. The analysis shows that the degeneracies in the low-redshift region in the $E(z)$ and $q(z)$ diagnostics are effectively broken by the statefinder diagnostic $r(z)$. By employing the statefinder diagnostic the holographic DE models can be differentiated efficiently in the low-redshift region; the degeneracy between the holographic DE models and the $\Lambda$CDM model can also be broken by this method. Especially for the HDE model, all the previous strong degeneracies are broken effectively. But for the NADE model, the degeneracy between the cases with different parameter values cannot be broken, even though the statefinder diagnostic is used.
Perhaps the statefinder hierarchy~\cite{Arabsalmani:2011fz} is helpful in breaking such a degeneracy in the
NADE model, and this possibility will be explored in a future work.
A direct comparison of the holographic DE models in the $r$--$s$ plane is also made. The current values $\{r_0,~s_0\}$ of the models play an important role in the statefinder analysis, since these values are thought to be extracted from the low-redshift data and the separations between the models (including the $\Lambda$CDM) can be directly measured in the $r$--$s$ plane. We hope that the future high-precision observations can offer more accurate data to discriminate DE models with the statefinder diagnostic and shed light on the nature of the dark energy.

\section*{Acknowledgements}
This work was supported by the National Nature Science Foundation of China (Grants No. 10975032 and No. 11175042), the National Ministry of Education of China (Grants No. NCET-09-0276 and No. N120505003) and the Education Department of Liaoning Province (Grants No. L2012087).

\end{document}